\documentclass[a4paper,11pt]{article}
\pdfoutput=1 
\usepackage{jheppub}
\usepackage[T1]{fontenc} \usepackage{comment}
\usepackage{amsmath, amsthm, amssymb, calrsfs, wasysym, verbatim, bbm, graphics, cancel}
\usepackage{tcolorbox}
\usepackage{mathrsfs}
\usepackage{physics}
\usepackage{slashed}
\usepackage{enumitem}
\usepackage{microtype}
\usepackage{hyperref}
\usepackage[nameinlink,noabbrev]{cleveref}
\usepackage{xcolor}
\usepackage[normalem]{ulem}

\usepackage{array}
\usepackage{tabularx}
\usepackage{booktabs}
\newcolumntype{Y}{>{\raggedright\arraybackslash}X}

\numberwithin{equation}{section}

\newcommand{\be}{\begin{equation}}
\newcommand{\ee}{\end{equation}}

\newcommand{\eqn}[1]{(\ref{#1})}

\title{\boldmath Gauge-covariant Hadamard descent in Yang–Mills theory}

\author[a,b]{L.~Vacchiano,}
\author[a,b]{G.~Chirco,}
\author[c,d]{S. Pascazio}
\author[a,b,e]{and P.~Vitale}

\affiliation[a]{Dipartimento di Fisica Ettore Pancini, Universit\`a degli Studi di Napoli Federico II,
Via Cintia 80126,  Napoli, Italy}
\affiliation[b]{INFN, Sezione di Napoli, Italy}
\affiliation[c]{Dipartimento di Fisica, Università di Bari, I-70126 Bari, Italy}
\affiliation[d]{INFN, Sezione di Bari, I-70125 Bari, Italy}
\affiliation[e]{School of Theoretical Physics Dublin Institute for Advanced Studies, 10 Burlington Road, Dublin 4, Ireland}

\emailAdd{lucio.vacchiano@unina.it}
\emailAdd{goffredo.chirco@unina.it}
\emailAdd{saverio.pascazio@uniba.it}
\emailAdd{patrizia.vitale@unina.it}

\begin{document}

\abstract{We study the dimensional reduction of  non-Abelian Yang-Mills theory from five to four dimensions. In this framework, Hadamard descent is formulated as a restriction to gauge-equivalence classes that are invariant, up to gauge transformations, under translations along the reduced direction.
{The central result is that, prior to gauge fixing, the adjoint scalar of the reduced theory is not the connection component $A_u$ itself, but the gauge-covariant combination $\widetilde{\Phi}=\epsilon-\iota_YA$, which follows naturally from the covariant Cartan identity. The usual identification with $A_u$ is recovered, up to the sign convention adopted here, in the $u$-independent gauge, while global holonomy data remain additional information to be specified separately.}
 We show that the reduced theory takes the form of a Yang--Mills--Higgs theory with the Higgs field in the adjoint representation. As an example, we focus on the $U(2)$ gauge theory and analyze the Higgs mechanism, together with the associated symmetry breaking pattern, when the reduced theory is expanded around the ground state of the parent theory. We then include fermions and show that, in the $u$-independent gauge, the extra-dimensional component of the gauge connection generates a pseudoscalar Yukawa coupling which, on the vacuum configuration, can be converted into a fermion mass term by a chiral field redefinition, while the gauge interaction remains vectorial.
}

\newcommand{\bb}[1]{\textcolor{blue}{#1}}
\maketitle

\flushbottom
\section{Introduction}

Dimensional reduction provides a natural bridge between field theories formulated in different spacetime dimensions. In the usual approach, one constructs a theory directly in the lower-dimensional spacetime, attempting to retain as many as possible of the characteristic properties of a higher-dimensional parent theory. A conceptually different question is to ask what becomes of the full higher-dimensional theory when the dependence on one spatial coordinate is suppressed. The lower-dimensional theory is then not introduced independently, but inherited from the parent one: field components that would otherwise be discarded may survive as additional degrees of freedom, while the original equations reorganize themselves into sectors reflecting the geometry and symmetries of the higher-dimensional theory.

This point of view is naturally associated with Hadamard’s method of descent. In his study of partial differential equations, Hadamard observed that a problem involving fewer variables may be obtained from one involving more variables by imposing uniformity along an additional coordinate~\cite{Hadamard}. Applied to field theory, the method amounts to restricting the higher-dimensional dynamics to configurations that are invariant, in an appropriate sense, along a chosen direction. The lower-dimensional theory is therefore obtained by decomposing the original fields rather than by postulating a new set of fields and interactions. All components compatible with the descent are retained, making explicit which structures are intrinsic to the lower-dimensional theory and which are inherited from the parent one.

Electromagnetism provides the simplest example. Electrodynamics in two spatial dimensions has been investigated from several points of view~\cite{lapidus,moreno,wheelerem,mcdonald,boito,GBB,BBG}. Under Hadamard descent, Maxwell theory separates into two lower-dimensional electromagnetic sectors, conventionally characterized by the field configurations EEB and BBE. These sectors are related by Hodge duality and, owing to the Abelian and linear character of Maxwell theory, remain dynamically decoupled. This perspective has been developed in previous studies of wave equations, electromagnetism, and Dirac fields~\cite{Huygens,ref-JMP,angelone23}.

The extension to non-Abelian gauge theories requires an important conceptual refinement. Whereas the electromagnetic field strength is gauge invariant, the Yang–Mills curvature transforms covariantly. Independence of the extra coordinate must therefore be formulated at the level of gauge-equivalence classes: translations along the redundant direction may change the gauge potential while leaving the physical configuration within the same gauge orbit. Moreover, because of the nonlinear commutator term in the curvature, field components that decouple in the Abelian case remain coupled after dimensional reduction.

Dimensional reduction and the emergence of adjoint scalar fields from higher-dimensional gauge connections are well-established results  \cite{ForgacsManton, Manton, Hosotani1, Hosotani2}. Related symmetry-based reductions of Yang–Mills theories have also been extensively developed in the framework of coset-space dimensional reduction \cite{KapetanakisZoupanos}. 
A dependence on the compact coordinate controlled by an internal-symmetry transformation also appears in Scherk–Schwarz reductions
\cite{ScherkSchwarz1979}, while extra-dimensional components of gauge connections play the role of Higgs fields and Wilson lines control symmetry breaking in gauge–Higgs unification \cite{CsakiGrojeanMurayama,Serone,HosotaniYamatsu}. The relation among three-dimensional Chern–Simons gravity, two-dimensional BF theories, and induced boundary dynamics has likewise been investigated in Refs. \cite{Ebert:2022ehb,ChircoVacchianoVitale2026,Mertens:2018fds}.



Our purpose here is more specific. We do not propose a new dimensional-reduction mechanism, nor do we claim that the emergence of adjoint scalar fields or the associated gauge-symmetry-breaking mechanism is new. Our contribution is to formulate Hadamard descent directly at the level of gauge-equivalence classes.
Let $Y=\partial_u$ generate translations along the direction to be reduced. Requiring such translations to be equivalent to an infinitesimal gauge transformation gives
$$
\mathcal{L}_Y A=D\epsilon .
$$
Together with the covariant Cartan identity, this condition singles out the combination
$$
\widetilde{\Phi}=\epsilon-\iota_YA, \qquad
\iota_YF=D\widetilde{\Phi},
$$
which transforms homogeneously in the adjoint representation and therefore provides the gauge-covariant scalar field of the reduced theory. The usual identification with the connection component $A_u$ is recovered, up to a sign convention, only after choosing the $u$-independent gauge $\epsilon=0$. In this way, the construction separates the local reduced dynamics from both the choice of a gauge representative and the global information encoded by holonomies. Unlike a Scherk–Schwarz reduction, no fixed global twist or monodromy is prescribed.

We apply this construction to the reduction of Yang–Mills theory from five to four spacetime dimensions. In the $u$-independent gauge, the five-dimensional action reduces to an ordinary four-dimensional Yang–Mills theory minimally coupled to an adjoint scalar. A constant ground-state configuration of this scalar selects a preferred direction in the gauge algebra: generators commuting with it form the unbroken subgroup and correspond to massless gauge fields, whereas the remaining gauge bosons acquire a mass. For the $U(2)$ model considered here, a generic non-central ground state produces the symmetry-breaking pattern
$$
U(2)\longrightarrow U(1)\times U(1),
$$
with two massless and two equally massive gauge bosons. The $U(2)$ example provides an explicit realization of the general centralizer construction and allows the mass spectrum to be determined directly.

The same higher-dimensional structure also controls the coupling to matter. When five-dimensional fermions are included, the covariant derivative along the reduced direction generates a four-dimensional pseudoscalar Yukawa coupling. On a constant ground-state configuration, a chiral field redefinition converts this coupling into an ordinary fermion mass term, while the gauge interaction remains vectorial. The adjoint scalar, the gauge-boson masses, and the fermion mass matrix thus have a common geometric origin in the extra-dimensional component of the gauge connection. Finally, the gauge-covariant formulation permits, at least locally along the reduced direction, the reconstruction of gauge orbits of solutions of the parent theory from solutions of the reduced equations.

The paper is organised as follows. In Section \ref{electro} we review the Hadamard descent method for electromagnetism in $D$ spacetime dimensions. In Section \ref{core} we formulate the non-Abelian reduction in a gauge-covariant way, specialize to the $u$-independent gauge, discuss the ground states and symmetry-breaking mechanism, and work out the $U(2)$ example. Section \ref{fermions} is devoted to the fermionic sector and the resulting Yukawa coupling and mass matrix. We summarize our results and discuss possible extensions in Section \ref{conclusions}.  The component formulation and the local reconstruction of the parent theory are presented in Appendix \ref {app:gauge-covariant-reduction}, while Appendix \ref{gaucovfermions} extends the gauge-covariant descent condition to fermionic fields.

\section{Dimensional Reduction of Electromagnetism}\label{electro}
In this section we briefly review the dimensional reduction of electromagnetism from $D$ to $D-1$ spacetime dimensions.\textcolor{black}{\footnote{Hadamard's method of descent  is applied in \cite{ref-JMP},  to the reduction of Maxwell theory from  $3+1$ to lower spatial dimensions.}} Consider Maxwell theory on a $D$-dimensional Minkowski spacetime with coordinates
\[
x^{I}=(x^{\mu},u), \qquad \mu = 0,\dots,D-2 ,
\]
where $u$ denotes the coordinate along which the reduction is performed. By ``redundant direction'' we mean that the physical object of the theory, namely the field strength $F_{IJ}$, does not depend on $u$:
\begin{equation}
\partial_{u}F_{IJ}=0 . \label{eq:1.1}
\end{equation}
If the manifold has trivial topology along $u$, this requirement can always be satisfied at the level of the potential, i.e.
\begin{equation}
\partial_{u}A_{I}=0. \label{eq:1.2}
\end{equation}
We then decompose the gauge potential as
\begin{equation}
A_{I}=(A_{\mu},\Phi), \qquad \Phi\equiv A_{u}, \label{eq:1.3}
\end{equation}
with $A_{\mu}$ and $\Phi$ independent of $u$ by \eqref{eq:1.2}. The field strength splits accordingly into a purely $(D-1)$-dimensional term and a mixed term,
\begin{equation}
F_{\mu\nu}=\partial_{\mu}A_{\nu}-\partial_{\nu}A_{\mu}, \qquad
F_{\mu u}=\partial_{\mu}\Phi-\partial_{u}A_{\mu}=\partial_{\mu}\Phi , \label{eq:1.4}
\end{equation}
the last equality following from \eqref{eq:1.2}.

Because the theory is Abelian, the $(D-1)$-dimensional Hodge dual is a genuine, well-defined operation on the reduced fields. Denoting by $\varepsilon_{\mu_{1}\dots\mu_{D-1}}$ the transverse Levi-Civita symbol, the equations of motion $\partial^{I}F_{IJ}=0$ decompose into a term for the free index $J=u$,
\begin{equation}
\partial^{\mu}F_{\mu u}=\partial^{\mu}\partial_{\mu}\Phi=0 , \label{eq:1.5a}
\end{equation}
and a term for the free index $J=\nu$,
\begin{equation}
\partial^{\mu}F_{\mu\nu}=0 , \label{eq:1.5b}
\end{equation}
together with the Bianchi identity $\partial_{[\mu}F_{\nu\rho]}=0$. Equations \eqref{eq:1.5a}--\eqref{eq:1.5b} are equivalent, component by component, under the exchange
\begin{equation}
F_{\mu\nu} \;\longleftrightarrow\; \varepsilon_{\mu\nu}{}^{\rho_{1}\cdots\rho_{D-3}}\,\partial_{\rho_{1}}\cdots(\star\,\partial\Phi)_{\rho_{2}\cdots\rho_{D-3}} ,
\end{equation}
i.e.\ the reduced field strength $F_{\mu\nu}$ is dual, in the transverse $(D-1)$-dimensional sense, to the gradient $\partial_{\mu}\Phi$ of the scalar $\Phi$. This is the component form of the $(D-1)$-dimensional 
Hodge duality between a one-form field strength and the derivative of a scalar. 
At the Lagrangian level we then have
\begin{equation}
    \mathcal{L}_{e.m.} =   F^{\mu \nu}F_{\mu \nu} + \partial^{\mu}\Phi \partial_{\mu}\Phi,
\end{equation}
therefore a dimensionally reduced electromagnetism is recovered in the $A_{\mu}$ sector together with a decoupled scalar field. The Abelian nature of Electromagnetism is responsible for the absence of any coupling between the two sectors. As we will see, this is not the case with non-Abelian gauge theories where the non linearity of the field strength implies a coupling between the reduced Yang-Mills theory and the additional scalar field.  

\section{
Gauge-covariant Hadamard descent in Yang--Mills theory}\label{core}
In this section we apply the dimensional reduction scheme for a non-Abelian gauge theory. 
\textcolor{black}{A methodological remark is in order. Hadamard's method of descent is most naturally and cleanly formulated at the level of the equations of motion and their solutions: one simply demands that the fields be uniform along the redundant direction, and the reduced equations follow directly. The situation becomes more involved when working at the level of the Lagrangian, since the latter is expressed in terms of potentials rather than field strengths, and additional assumptions are required to close the procedure. In the non-Abelian case this difficulty is compounded: unlike the Abelian setting, the potentials appear explicitly already in the equations of motion, through the nonlinear commutator terms in the curvature. A well-defined reduction therefore requires a precise transformation prescription for the non-Abelian potentials under descent. Specifying such a prescription is a necessary working assumption of our approach. 
}
\\ \\
For a non-Abelian gauge theory the curvature two-form is not gauge invariant, rather it transforms 
as:
\begin{equation}
    F_{\mu \nu} \rightarrow g^{-1}F_{\mu \nu}g,
\end{equation}
therefore, the $u$-independence of the field strength is not a gauge  invariant statement. Second, the non-linearity of Yang-Mills theory implies that the reduced theory describes two interacting fields $A^a_{\mu}$ and $\Phi^a$. 
These are the main differences from the Abelian Hadamard descent, where the two reduced sectors are completely decoupled.
We are going to explore these differences in what follows.

\subsection{Gauge-covariant reduction}
We keep the same index conventions as above, with the addition of a Lie-algebra (``color'') index $a,b,c,\dots$ running over a basis $\tau_{a}$ of the gauge algebra, normalized so that
\begin{equation}
\tau_{a}^{\dagger}=-\tau_{a}, \qquad \exp(\theta^{a}\tau_{a})\in G \ \text{for real } \theta^{a}, \qquad [\tau_{b},\tau_{c}]=f^{a}{}_{bc}\,\tau_{a}.
\end{equation}
A good notion of dimensional reduction for a Yang--Mills field consists in requiring that any dependence of the gauge potential $A_{I}^{a}$ on $u$ can be undone by a (possibly $u$-dependent) gauge transformation, where $u$ is a coordinate along the redundant direction.  On a $D$-dimensional manifold $M_{D}$, the field strength components 
\begin{equation}
F^{a}_{IJ}=\partial_{I}A^{a}_{J}-\partial_{J}A^{a}_{I}+f^{a}{}_{bc}A^{b}_{I}A^{c}_{J} \label{eq:5.1a}
\end{equation}
transform under a  local gauge transformation $g(x)\in G$,  in the adjoint representation.
Let $Y=\partial_{u}$ generate translations along the reduced direction. Demanding that the physics is independent of $u$ means that translations along $u$ are equivalent to a gauge transformation,
\begin{equation}
F^{a}_{IJ}(x,u)=\big(\mathrm{Ad}_{g^{-1}(x,u)}\big)^{a}{}_{b}\,F^{b}_{IJ}(x). \label{eq:5.2}
\end{equation}
At the level of the potential, this ``covariant constancy'' condition reads
\be \mathcal{ L}_YA=D\epsilon
\ee
namely,
\begin{equation}
\partial_{u}A^{a}_{I}=D_{I}\epsilon^a\equiv \partial_{I}\epsilon^a+f^{a}{}_{bc}A^{b}_{I}\epsilon^{c}, \label{eq:5.3}
\end{equation}
where $\epsilon^a(x,u)$ is a Lie-algebra-valued parameter associated with $Y$. Consistency is checked by evolving $F^{a}_{IJ}$ along $Y$: using \eqref{eq:5.1a} and \eqref{eq:5.3},
\begin{equation}
\partial_{u}F^{a}_{IJ}(x,u)=D_{I}\big(D_{J}\epsilon^a\big)-D_{J}\big(D_{I}\epsilon^a\big)=f^{a}{}_{bc}F^{b}_{IJ}(x,u)\epsilon^{c}(x,u), \label{eq:5.4}
\end{equation}
which corresponds, in index free notation, to   $\mathcal{L}_{Y}F=[F,\epsilon]$: the field strength remains within the same gauge orbit along the reduced direction. Eq.  \eqn{eq:5.4} is  the non-Abelian condition for dimensional reduction. 
The requirement that spacetime transformations be compensated by
gauge transformations is closely related to the equivariance
condition appearing in coset-space dimensional reduction
\cite{ForgacsManton, KapetanakisZoupanos}. In the present setting, however, this condition is
interpreted as the gauge-covariant implementation of Hadamard
descent along a single redundant direction
and generalizes the descent condition of the Abelian case \cite{ref-JMP,angelone23}.

Using the covariant Cartan identity
\be
\mathcal{L}_Y A=\iota_Y F+D(\iota_Y A) 
\ee
{with the following convention for the interior product, $i_Y$,
$(\iota_Y F)(X)=F(Y,X)$
[hence, for $Y=\partial_u$,
$(\iota_YF)_\mu=F_{u\mu}=-F_{\mu u}$], }
one  obtains 
\be
\iota_Y F =D\left(\epsilon-\iota_YA\right).
\ee
On defining 
$\widetilde\Phi\equiv\epsilon-\iota_Y A, $
we arrive at 
\be \iota_Y F=D\widetilde\Phi.\label{nonabDR}
\ee
This equation is the gauge-covariant non-Abelian counterpart of
the Abelian relation $F_{\mu u}=\partial_\mu\Phi$.

The compensating field $\epsilon$  is not gauge covariant by itself. Under a gauge transformation $h(x,u)$, both $\epsilon$ and $\Phi=A_u$ acquire the same inhomogeneous term $h^{-1}\partial_u h$. Their difference $\widetilde\Phi=\epsilon-\Phi$ therefore transforms homogeneously in the adjoint representation. It is this covariant combination that plays the role of the scalar field in the reduced equations. 

A detailed component derivation of the reduced equations, together
with the local reconstruction of the parent solutions, is given in
Appendix~\ref{app:gauge-covariant-reduction}.

\subsection{The $u$-independent gauge}
From now on we are going to fix $\epsilon^a(x,u)=0$, which means $F^a_{IJ}=F^a_{IJ}(x)$. Consistently, the associated gauge potential $A^a_J$ is $u-$independent and $\widetilde\Phi= -\Phi$.  The $D$-dimensional connection is therefore decomposed into a $(D-1)$-dimensional gauge field $A^{a}_{\mu}$ and a scalar $\Phi^{a}$ (the component along $du$),
\begin{equation}
A^{a}_{I}=(A^{a}_{\mu},\Phi^{a}), \qquad \Phi^{a}\equiv A^{a}_{u}, \label{eq:2.1}
\end{equation}
with $A^{a}_{\mu}, \Phi^{a}$ independent on $u$. The reduced field strength reads
\begin{align}
F^{a}_{\mu\nu}&=\partial_{\mu}A^{a}_{\nu}-\partial_{\nu}A^{a}_{\mu}+f^{a}{}_{bc}A^{b}_{\mu}A^{c}_{\nu}, \label{eq:2.2a}\\
F^{a}_{\mu u}&\equiv D_{\mu}\Phi^{a}=\partial_{\mu}\Phi^{a}+f^{a}{}_{bc}A^{b}_{\mu}\Phi^{c}, \label{eq:2.2b}
\end{align}
where $D_{\mu}$ denotes the covariant derivative with respect to $A^{a}_{\mu}$ in the adjoint representation. 
The equations of motion
\begin{equation}
    \partial^I F^a_{IJ} + f^a_{bc}A^{b,I}F^c_{IJ}=0,
\end{equation}
reduce to
\begin{equation}
\label{eq.red.F}
    D^\mu F^a_{\mu \nu} + f^a_{bc}\,\Phi^b D_{\nu}\Phi^c=0,
\end{equation}
\begin{equation}
\label{eq.red.Phi}
    D^\mu D_{\mu}\Phi^a =0.
\end{equation}
These are the equations of motion of a Yang-Mills theory with a minimally coupled Higgs field in the adjoint representation. They can be recovered by also performing a dimensional reduction of the $D$-dimensional action for the pure Yang-Mills theory as follows. Let $K_{ab}$ be the components of an invariant, non-degenerate bilinear form on the Lie algebra (replacing the trace pairing $\Tr(\cdot\,,\cdot)$ of the index-free formulation). The $D$-dimensional Yang--Mills action reduces as
\begin{equation}
S_{YM}=\textcolor{black}{-\frac{1}{4} \int_{M_{D-1}\times S^{1}} K_{ab}\,F^{a}_{IJ}F^{b,IJ}
=2\pi R\int_{M_{D-1}} \Big( -\frac{1}{4} K_{ab}F^{a}_{\mu\nu}F^{b,\mu\nu}+\frac{1}{2} K_{ab} D_{\mu}\Phi^{a}D^{\mu}\Phi^{b}\Big)}, \label{eq:2.3}
\end{equation}
whose equations of motion are Eqs.\  \eqn{eq.red.F} and \eqn{eq.red.Phi}. This shows that dimensional reduction naturally generates a \textbf{Yang--Mills--Higgs} theory: $\Phi^{a}$ transforms in the adjoint representation and plays the role of a Higgs field. If $\Phi^{a}$ acquires a vacuum expectation value $H^{a}$, the Higgs mechanism is triggered and a subset of the gauge bosons $A^{a}_{\mu}$ becomes massive. In particular, the gauge fields whose generators commute with $H=H^{a}\tau_{a}$ remain massless, while all others acquire a mass.
\subsection{Ground states and residual gauge symmetry}

A ground state is a  field configuration that minimizes the energy; for a pure Yang--Mills theory in components this is the flat-connection condition
\begin{equation}
F^{a}_{IJ}=0 . \label{eq:2.4}
\end{equation}
 Dimensional reduction introduces a major simplification: since the vacuum condition also involves $\Phi^{a}$ through $F^{a}_{\mu u}=D_{\mu}\Phi^{a}$, locally the vacuum configurations  are parametrized by the constant adjoint
scalar $\Phi^a=H^a$ alone, after gauging the flat transverse connection to zero. Globally, additional data may arise from non-trivial flat transverse connections.
  Under dimensional reduction, $\Phi^{a}$ ceases to transform as a connection component and transforms instead as a vector in the adjoint representation, while $A^{a}_{\mu}$ continues to transform as a genuine connection. On a vacuum configuration,
\begin{equation}
F_{\mu\nu}^{a}\big|_{0}=0, \qquad D_{\mu}\Phi^{a}\big|_{0}=0. \label{eq:2.5}
\end{equation}
Locally, $A^{a}_{\mu}$ can always be gauged to zero, but the same is not true of $\Phi^{a}$; setting $A^{a}_{\mu}=0$ in \eqref{eq:2.5} gives
\begin{equation}
\partial_{\mu}\Phi^{a}\big|_{0}=0, \label{eq:2.6}
\end{equation}
so vacuum configurations are constant, $\Phi^{a}=H^{a}$, on the transverse manifold $M_{\perp}$. If $M_{D-1}$ (equivalently $M_{\perp}$) is not simply connected, $A^{a}_{\mu}$ cannot be set to zero globally, and more exotic vacuum configurations become possible. 
The scalar $\Phi^{a}$ thus fixes the vacuum of the theory without the need for an additional potential term. 
\\
\textcolor{black}{
In the full
higher-dimensional theory, a constant $H$ can be locally removed by
the gauge transformation
\[
g(u)=\exp(-uH).
\]
If the extra dimension is a circle of radius $R$,  this transformation is
single-valued only if
\[
\exp(2\pi R H)=\mathbbm{1},
\]
namely if the corresponding holonomy is trivial. In this case the
background $A_u=H$ is globally gauge-equivalent to $A_u=0$. However, the transformation
$g(u)=\exp(-uH)$ does not in general preserve the reduced sector:
for a generic $u$-independent transverse gauge field one has
\[
A_\mu(x)\longmapsto
A_\mu^g(x,u)=g^{-1}(u)A_\mu(x)g(u),
\]
which depends on $u$ whenever $[A_\mu,H]\neq0$. It is therefore not
an allowed residual gauge transformation of the dimensionally
reduced theory, whose gauge transformations are required to preserve
$u$-independence. Within the reduced theory, $H$ can only be
conjugated by $u$-independent gauge transformations, and its
centralizer determines the unbroken gauge group. 
The full higher-dimensional theory also admits flat connections
with non-trivial holonomy. Since the analysis below concerns the
local reduced dynamics, we do not select a particular holonomy
sector. Global restrictions on $H$ must be imposed separately,
according to the topology and the boundary conditions along the
extra dimension.}

\subsection{The $U(2)$ example}

For a $U(2)$ gauge theory we take the basis
\begin{equation}\label{basis}
\tau_{a}=(iI,\,i\sigma_{1},\,i\sigma_{2},\,i\sigma_{3}), \qquad a=0,1,2,3,
\end{equation}
with $\sigma_{i}$ the Pauli matrices, corresponding to the algebra decomposition $\mathfrak{u}(2)=\mathfrak{u}(1)\oplus \mathfrak{su}(2)$. The most general invariant bilinear form compatible with this splitting is
\begin{equation}
K_{ab}=-k_1\,\Tr(\tau_{a}\tau_{b})-k_2\,\Tr(\tau_{a})\Tr(\tau_{b}), \label{eq:2.8}
\end{equation}
so that $k_1$ is the coupling of the $\mathfrak{su}(2)$ sector and $k_2$ that of the $\mathfrak{u}(1)$ sector. In the basis~\eqn{basis}, the invariant form is block diagonal:
$K_{ij}= 2 k_1\delta_{ij}$ on the $\mathfrak{su}(2)$ sector, whereas
$K_{00}$ receives contributions from both coefficients $k_1$
and $k_2$. Equivalently, the two independent parameters determine
the relative normalizations of the central $\mathfrak{u}(1)$ and $\mathfrak{su}(2)$
sectors. The gauge coupling of the fermions can, in principle, be chosen independently of $k_1,k_2$.

\subsection{Gauge-boson masses}\label{masses}
From this point on, we restore explicitly the gauge coupling $g$,
so that the adjoint covariant derivative reads
$D_\mu H=\partial_\mu H+g[A_\mu,H]$.
Gauge bosons acquire a mass when $\Phi^{a}$ takes a vacuum expectation value $H^{a}$.   From  Eqs. \eqref{eq:2.2b}, \eqn{eq:2.6},
\begin{equation}
D_{\mu}H^{a}=g f^{a}{}_{bc}A^{b}_{\mu}H^{c}, \label{eq:2.9}
\end{equation}
and squaring this expression (contracted with $K_{ab}$) yields the mass term
\begin{equation}
\mathcal{L}_{\rm mass}=\frac{g^2}{2} K_{ab}\big(f^{a}{}_{mn}A^{m}_{\mu}H^{n}\big)\big(f^{b}{}_{pq}A^{p,\mu}H^{q}\big). \label{eq:2.10}
\end{equation}
Hence every component $A^{b}_{\mu}$ for which $f^{a}{}_{bc}A^{b}_{\mu}H^{c}$ does not vanish identically becomes massive: the gauge bosons split into those whose generators commute with $H$ (massless) and those that do not (massive). Classifying the massive directions amounts to finding the centralizer of $H$ modulo gauge transformations.

Let us write
\begin{equation}
H^{a}=(k,\,\vec h) , \qquad H=i\big(k\,I+\vec h\cdot\vec\sigma\big). \label{eq:2.11}
\end{equation}
Since $H^{0}=k$ multiplies the central generator $\tau_{0}=iI$, it never contributes to \eqref{eq:2.10}, because $f^{a}{}_{bc}=0$ whenever an index equals $0$. Only the $\mathfrak{su}(2)$ term, $\vec h$, matters. Using the $SU(2)$--$SO(3)$ homomorphism, conjugation of $H$ by $g\in SU(2)$ acts on $\vec h$ as an $SO(3)$ rotation,
\begin{equation}
\vec h \;\longrightarrow\; R_{g}\vec h , \label{eq:2.12}
\end{equation}
so that, $g$ being arbitrary, $\vec h$ can always be rotated into the standard form $\vec h=(0,0,{\beta} )$. We thus set, without loss of generality,
\begin{equation}
H^{0}={\alpha}, \qquad H^{3}=\beta, \qquad H^{1}=H^{2}=0. \label{eq:2.13}
\end{equation}
\textcolor{black}{Using
\[
[\tau_i,\tau_j]=-2\epsilon_{ijk}\tau_k,
\]
we obtain
\be
D_\mu H
 =g[A_\mu,H]
 =2g\beta\left(A_\mu^1\tau_2-A_\mu^2\tau_1\right).
\ee
Therefore, for $K_{ij}=2k_1\delta_{ij}$, the mass contribution is
\be \label{eq:2.14}
\mathcal{L}_{\mathrm{mass}}
 =4k_1 g^2\beta^2
 \left(A_\mu^1A^{1\mu}+A_\mu^2A^{2\mu}\right).
\ee
After canonical normalization of the gauge fields, the two massive
gauge bosons have equal mass
\be
m_1^2=m_2^2=4g^2\beta^2,
\ee
whereas $A_\mu^0$ and $A_\mu^3$ remain massless.
}
There are therefore only two possible scenarios: if $\beta=0$ no gauge boson is massive; if $\beta\neq0$ exactly $A^{1}_{\mu},A^{2}_{\mu}$ become massive, with $m^{2}=4 g^{2}\beta^{2}$. The only allowed symmetry-breaking pattern is
\begin{equation}
U(2)\;\longrightarrow\;U(1)\times U(1).
\end{equation}
As will be seen in the fermionic sector below, even when $\beta=0$ the two flavor components acquire the same mass while the gauge bosons stay massless; when $\beta=\pm\alpha$, only one flavor component acquires a mass, but in this case two gauge bosons are necessarily massive.

\section{Fermionic Sector and Yukawa Couplings}\label{fermions}

In order to study the fermionic sector of the reduced theory and the mass generation,  we consider in  this Section  the Dirac equation in 5 dimensions with $U(2)$ coupling.  Fermion masses and Yukawa-type interactions induced by the extra-dimensional components of gauge connections have also been studied in the context of gauge-Higgs unification \cite{ScruccaSeroneSilvestrini}. The free Dirac equation within the Hadamard descent method was already considered in \cite{angelone23}.

To include fermions we work in $5$ dimensions with signature $(+,-,-,-,-)$ and coordinates $(t,x,y,z,u)$. The $5$D gamma matrices $\Gamma^{I}$ satisfy $\{\Gamma^{I},\Gamma^{J}\}=2\eta^{IJ}$ and can be represented using the $4$D Dirac matrices $\gamma^{\mu}$ together with
\begin{equation}
\Gamma^{\mu}=\gamma^{\mu} \ (\mu=0,1,2,3), \qquad \Gamma^{4}=i\gamma^{5}. \label{eq:3.1}
\end{equation}
The $5$D Dirac action for a fermion $\Psi$ coupled to $A^{a}_{I}$ is
\begin{equation}
S_{F}=\int_{M_{5}} \bar\Psi\big(i\Gamma^{I}D_{I}\big)\Psi\, d^{5}x
=\int_{M_{5}} \bar\Psi\big(i\gamma^{\mu}D_{\mu}+i\Gamma^{4}D_{u}\big)\Psi\,d^{5}x, \label{eq:3.2}
\end{equation}
with
\begin{equation}
D_{\mu}=\partial_{\mu}+g\,A^{a}_{\mu}\tau_{a}, \qquad D_{u}=\partial_{u}+g\,\Phi^{a}\tau_{a}=g\,\Phi^{a}\tau_{a}, \label{eq:3.4+3.6}
\end{equation}
the last equality holding once the zero-mode condition $\partial_{u}\Psi=0$ is imposed, together with the identification $A^{a}_{u}=\Phi^{a}$ of \eqref{eq:2.1}. The condition $\partial_u \Psi =0$ can  also be promoted to a covariant one $\partial_u \Psi = - g\epsilon \Psi$, which realizes a Yukawa coupling with the aforementioned $\tilde{\Phi}$  \textcolor{black}{(see Appendix \ref{gaucovfermions} for details)}.

When dealing with fermions, the number of initial spacetime dimensions plays a crucial role since the dimension of the Clifford representation doubles when the spacetime dimension increases by  two. \textcolor{black}{ A Dirac spinor in five dimensions has four complex components and, when viewed from the four-dimensional Lorentz subgroup, decomposes into two Weyl spinors. In five dimensions, however, there is no intrinsic chirality operator.} The difference with four-dimensional fermions is in the $\gamma^5$, which in 5D is a genuine generator of the Clifford algebra, therefore the notion of chirality is lost. We will see that this implies that gauge fields couple with \textcolor{black}{vector currents} associated with mass eigenstates.   
\subsection{Pseudoscalar Yukawa coupling and fermion masses}

Using \eqref{eq:3.1} and \eqref{eq:3.4+3.6}, the term $i\Gamma^{4}D_{u}$ in \eqref{eq:3.2} becomes
\begin{equation}
i(i\gamma^{5})\,g\,\Phi^{a}\tau_{a}=-g\,\gamma^{5}\Phi^{a}\tau_{a}. \label{eq:3.5}
\end{equation}
The effective $4$D action is
\begin{equation}
S_{F,4D}=2\pi R\int_{M_{4}} \bar\Psi\Big(i\gamma^{\mu}D_{\mu}-g\,\gamma^{5}\Phi^{a}\tau_{a}\Big)\Psi\, d^{4}x. \label{eq:3.6b}
\end{equation}
The term $-g\,\bar\Psi\gamma^{5}\Phi^{a}\tau_{a}\Psi$ is a \textcolor{black}{pseudo-scalar } Yukawa coupling; along the vacuum expectation value $\Phi^{a}\to H^{a}$ it reduces to $-g\bar\Psi\gamma^{5}H\Psi$ with $H=\,H^{a}\tau_{a}$. This term is odd under parity, since $\Psi(t,\vec x)\to\gamma^{0}\Psi(t,-\vec x)$ and $\gamma^{0}\gamma^{5}\gamma^{0}=-\gamma^{5}$, and it is not directly a mass term because of $\gamma^{5}$. A genuine mass term is obtained after the chiral field redefinition
\begin{equation}
\Psi=e^{i\theta\gamma^{5}}\chi, \qquad \bar\Psi=\bar\chi\,e^{i\theta\gamma^{5}}, \label{eq:3.7-3.9}
\end{equation}
using $\{\gamma^{5},\gamma^{0}\}=0$. Then
\begin{equation}
-g \bar\Psi\gamma^{5}H\Psi=-g \bar\chi\,\gamma^{5}e^{2i\theta\gamma^{5}}H\chi,
\qquad e^{2i\theta\gamma^{5}}=\cos(2\theta)+i\gamma^{5}\sin(2\theta), \label{eq:3.10-3.11}
\end{equation}
using $(\gamma^{5})^{2}=I$. Choosing $\theta=\pi/4$,
\begin{equation}
-g \bar\Psi\gamma^{5}H\Psi = - \bar\chi\,(ig H)\,\chi. \label{eq:3.12}
\end{equation}
Since $iH$ is Hermitian, $M\equiv ig H$ is a genuine, diagonalizable mass matrix with real eigenvalues; the fermions $\chi$ are the mass eigenstates. \textcolor{black}{This conclusion holds at the classical level. At the quantum level,
the chiral field redefinition \eqn{eq:3.7-3.9} may induce a non-trivial Jacobian
in the fermionic functional measure and consequently generate a
topological contribution through the chiral anomaly
\cite{Fujikawa,FujikawaSuzuki}. In the present analysis we restrict
ourselves to the classical reduced theory. }

However, rewriting the gauge interaction in terms of $\chi$ shows that the interaction term does not change, in fact the interaction term $\bar{\Psi}\gamma^\mu A^a_{\mu}\tau_a\Psi$ becomes
\begin{equation}
   \mathcal{L}_{int}=\, i g \bar{\chi}\gamma^{\mu}A_{\mu}^a \tau_a \chi.
\end{equation}
In the $U(2)$ case one may need in addition a genuine flavor rotation $U$ diagonalizing $M$,
\begin{equation}
M_{\rm diag}=\mathrm{diag}(m_{1},m_{2})=U^{-1}(ig H)U, \label{eq:3.13}
\end{equation}
which correspondingly rotates the generators appearing in the covariant derivative,
\begin{equation}
\partial_{\mu}+g A^{a}_{\mu}\tau_{a} \;\longrightarrow\; \partial_{\mu}+g A^{a}_{\mu}\,U^{-1}\tau_{a}U, \label{eq:3.14}
\end{equation}
in close analogy with the fermion mixing familiar from electroweak theory.

Writing $\chi$ as a $U(2)$ doublet $\chi=(\chi_{a},\chi_{b})^{T}$ and using $H=i(\alpha I+\beta\sigma_{3})$ from \eqref{eq:2.13}, so that $-iH=\alpha I+\beta\sigma_{3}$,
\begin{equation}
-\bar\chi\,(ig H)\,\chi=g \bar\chi\begin{pmatrix}\alpha+\beta & 0\\ 0 & \alpha-\beta\end{pmatrix}\begin{pmatrix}\chi_{a}\\ \chi_{b}\end{pmatrix}. \label{eq:3.15}
\end{equation}
By tuning $\alpha,\beta$ one may obtain degenerate eigenvalues (equal masses for both flavors, as when $\beta=0$), or, choosing $\alpha=\pm\beta$, one eigenvalue vanishes while the other does not, so that only one flavor component acquires a mass -- consistently with the gauge-boson analysis of Section \ref{masses}.

The gauge-boson and fermion spectra for the different values of
$\alpha$ and $\beta$ are summarized in
Table~\ref{tab:fermion-gauge-masses}.

\begin{table}[htbp]
\centering
\footnotesize
\renewcommand{\arraystretch}{1.25}

\begin{tabularx}{\textwidth}{@{}
  >{\raggedright\arraybackslash}p{0.15\textwidth}
  Y
  Y
  >{\raggedright\arraybackslash}p{0.18\textwidth}
  @{}}

\toprule
\textbf{Case}
&
\textbf{Gauge-boson masses}
&
\textbf{Fermion masses}
&
\textbf{Residual group}
\\
\midrule

$\beta=0$
&
$m_{A^a}^2=0$, $a=0,1,2,3$
&
$m_1=m_2=g|\alpha|$
\newline
(both vanish for $\alpha=0$)
&
$U(2)$
\\
\addlinespace

$\alpha=\beta\neq0$
&
$m_{A^1}^2=m_{A^2}^2=4g^2\beta^2$
\newline
$m_{A^0}^2=m_{A^3}^2=0$
&
$m_1=2g|\beta|$
\newline
$m_2=0$
&
$U(1)\times U(1)$
\\
\addlinespace

$\alpha=-\beta\neq0$
&
$m_{A^1}^2=m_{A^2}^2=4g^2\beta^2$
\newline
$m_{A^0}^2=m_{A^3}^2=0$
&
$m_1=0$
\newline
$m_2=2g|\beta|$
&
$U(1)\times U(1)$
\\
\addlinespace

$\beta\neq0$,
\newline
$\alpha\neq\pm\beta$
&
$m_{A^1}^2=m_{A^2}^2=4g^2\beta^2$
\newline
$m_{A^0}^2=m_{A^3}^2=0$
&
$m_1=g|\alpha+\beta|$
\newline
$m_2=g|\alpha-\beta|$
\newline
(degenerate only for $\alpha=0$)
&
$U(1)\times U(1)$
\\

\bottomrule
\end{tabularx}

\caption{Gauge-boson spectrum, fermion masses, and residual gauge
symmetry in the $U(2)$ model. Gauge-symmetry breaking depends only
on $\beta$, whereas the fermion spectrum depends on both $\alpha$
and $\beta$. For $\beta\neq0$, $A_\mu^1$ and $A_\mu^2$ are massive,
while $A_\mu^0$ and $A_\mu^3$ remain massless. For
$\alpha=\pm\beta\neq0$, one fermion eigenstate is massless.}

\label{tab:fermion-gauge-masses}
\end{table}

\section{Conclusions}\label{conclusions}
In this work we have formulated the dimensional reduction of non-Abelian Yang--Mills theory as a gauge-covariant extension of Hadamard descent. Rather than requiring the gauge potential itself to be independent of the redundant coordinate, we require its variation along the corresponding vector field $Y$ to be a gauge transformation,
\[
\mathcal{L}_Y A=D\epsilon .
\]
Combining this condition with the covariant Cartan identity leads naturally to the adjoint field
\[
\widetilde{\Phi}=\epsilon-\iota_YA,
\qquad
\iota_YF=D\widetilde{\Phi}.
\]
The inhomogeneous terms in the gauge transformations of $\epsilon$ and $\iota_YA$ cancel, so that $\widetilde{\Phi}$ transforms homogeneously in the adjoint representation. This identifies the gauge-covariant object that plays the role of the scalar field in the reduced theory.
{The main outcome is therefore a gauge-independent characterization of Hadamard descent: the reduced scalar is defined intrinsically on gauge orbits, whereas its identification with an extra-dimensional component of the connection arises only after choosing a $u$-independent representative. This also makes explicit which aspects of the reduction are local and which depend on global holonomy data.}

In the $u$-independent gauge $\epsilon=0$, the construction reduces to the usual decomposition of the higher-dimensional connection into a lower-dimensional gauge field and an adjoint scalar. The higher-dimensional Yang--Mills equations and action then give a Yang--Mills theory minimally coupled to this scalar. A constant non-central vacuum configuration $H$ breaks the residual gauge symmetry to the centralizer of $H$. For the $U(2)$ model considered here, a generic non-central background produces the symmetry-breaking pattern
$
U(2)\longrightarrow U(1)\times U(1),
$
with two massless and two equally massive gauge bosons. With the generator normalization adopted in this work, the latter have
\[
m_1^2=m_2^2=4g^2\beta^2.
\]
We have also considered the reduction of the fermionic sector. The component of the higher-dimensional covariant derivative along the reduced direction gives rise to a pseudoscalar Yukawa coupling. On a constant vacuum configuration, this coupling can be converted classically into an ordinary fermion mass term by a chiral field redefinition, while the gauge interaction remains vectorial. At the quantum level, the same redefinition may induce a non-trivial Jacobian in the fermionic measure through the chiral anomaly; the present analysis has been restricted to the classical reduced theory. A full quantum mechanical analysis of the Hadamard descent is an  interesting open problem to be addressed in future works.
The gauge-covariant formulation developed here separates the local content of Hadamard descent from the global information carried by holonomies and boundary conditions. It may therefore provide a useful starting point for extending the construction to non-trivial bundles, more general gauge systems and compactifications. A particularly interesting direction is the reduction from four to three dimensions, where the fermionic sector and its induced mass structure may have applications to planar and condensed-matter systems.

\section*{Acknowledgments}
G.C., L.V. and P.V. acknowledge support from the INFN Iniziativa Specifica GeoSymQFT. They also acknowledge support from the European COST Actions   BridgeQG  CA23130 and RQI  CA23115.  L.V. thanks the group of G. Barnich and G. Compère for fruitful discussions and the Theoretical and Mathematical Physics of the ULB for hospitality while this work was being completed. 
S.P. acknowledges support from INFN through the project ``QUANTUM", from the Italian funding within the ``Budget MUR - Dipartimenti di Eccellenza 2023–2027" - Quantum Sensing and Modelling for One-Health (QuaSiModO),  from Projects PN-RIC Q-SUD (Q-SUD B99H26000380007) and PN-RIC QUANTAS (SYNERGIA B99H26000410007).

\appendix
\section{Component formulation and local reconstruction}
\label{app:gauge-covariant-reduction}
In order to perform the dimensional reduction in a gauge covariant way, let us  consider  the {infinitesimal gauge transformation along the $u$ direction, as in \eqn{eq:5.3} }
\begin{equation}\label{gatra}
    \partial_u A_{\mu} = [A_{\mu}, \epsilon] + \partial_{\mu}\epsilon, \quad     \partial_u \Phi = [\Phi, \epsilon] + \partial_u \epsilon
\end{equation}
 Then the field strength components
\begin{equation}
    F_{u\mu} = \partial_u A_{\mu}- \partial_{\mu}\Phi + [\Phi, A_{\mu}]
\end{equation}
become 
\begin{equation}
    F_{u \mu} = D_{\mu}\tilde{\Phi},
\end{equation}
where $\tilde{\Phi} = \epsilon - \Phi$. {By applying the DR-prescription \eqn{eq:5.4} we also know that }
\begin{equation}
\partial_u F_{u \mu} = [F_{u \mu}, \epsilon], \quad \partial_u F_{\mu \nu} = [F_{\mu \nu}, \epsilon].
\end{equation}
{ The equations of motion $D^I F_{IJ}=0$ for $J=\mu$, namely  
\begin{equation}
    \partial^u F_{u \mu} + [\Phi, F_{u \mu}] + D^{\nu}F_{\nu \mu} =0,
\end{equation}
then become
\begin{equation} \label{A6}
    [D_{\mu}\tilde{\Phi}, \tilde{\Phi}] + D^{\nu}F_{\nu \mu} =0.
\end{equation}
while posing $J=u$, namely, 
\be
D^\nu F_{\nu u}=0
\ee
we get} 
\begin{equation}\label{neweq}
    D^{\nu}D_{\nu}\tilde{\Phi}=0.
\end{equation}
{The compensating field $\epsilon$  is not gauge covariant by itself. Under a gauge transformation $h(x,u)$, both $\epsilon$ and $\Phi=A_u$ transform inhomogeneously
\begin{eqnarray}
A^h&=&h^{-1}Ah+h^{-1}dh, \nonumber \\
 \epsilon^h &=&h^{-1}\epsilon h+h^{-1}\partial_u h .
 \end{eqnarray}
Since 
\be
(A_u)^h =h^{-1}(A_u)h+ h^{-1}\partial_u h, 
\ee
the inhomogeneous terms cancel out to give 
\be
\widetilde\Phi^h =h^{-1}\widetilde\Phi h. 
\ee
Therefore, 
$\widetilde\Phi$ may be interpreted as an adjoint scalar field. 
 The equations of motion are the same as those of the reduced theory but with the full $u$-dependence restored.    
Equations~\eqn{A6}   and~\eqn{neweq}  contain no explicit derivatives along the
$u$ direction. At every fixed value $u=u_0$, they therefore take the
form of a Yang--Mills theory for $A_\mu(x,u_0)$ minimally coupled to
the adjoint scalar $\widetilde{\Phi}(x,u_0)$. Thus, locally along the
reduced direction, every solution of the reduced equations determines
a gauge orbit of solutions of the parent theory. Global reconstruction
may additionally require compatibility with the periodicity or
boundary conditions imposed along the $u$ direction.}

\section{Gauge-covariant descent of fermionic fields}\label{gaucovfermions}
When fermions are minimally coupled, the gauge-covariant descent
condition reads
\begin{equation}
    \partial_u\Psi=-g\epsilon^a\tau_a\Psi .
\end{equation}
Indeed,
\begin{equation}
    D_u\Psi
    =g(\Phi^a-\epsilon^a)\tau_a\Psi
    =-g\widetilde{\Phi}\Psi ,
\end{equation}
and therefore, using $\Gamma^4=i\gamma^5$,
\begin{equation}
    i\Gamma^4D_u\Psi
    =g\gamma^5(\epsilon^a-\Phi^a)\tau_a\Psi
    =g\gamma^5\widetilde{\Phi}\Psi .
\end{equation}
In the $u$-independent gauge $\epsilon=0$, this reduces to the
pseudoscalar Yukawa coupling obtained in Eq.~\eqn{eq:3.5}.

\appendix


\begin{thebibliography}{99}

\bibitem{Hadamard}
J.~Hadamard,
\emph{Lectures on Cauchy's problem in linear partial differential equations},
Yale University Press, New Haven (1923).

\bibitem{lapidus}
I.R.~Lapidus,
\emph{Classical electrodynamics in a universe with two space dimensions},
\textit{Am. J. Phys.} \textbf{50} (1982) 155.

\bibitem{moreno}
E.~Moreno and J.~Rivas,
\emph{On electromagnetism in $\mathbb{R}^{2+1}$},
\textit{Eur. J. Phys.} \textbf{5} (1984) 20.

\bibitem{wheelerem}
N.~Wheeler,
\emph{``Electrodynamics'' in 2-dimensional spacetime},
Reed College (1997),
\url{https://www.reed.edu/physics/faculty/wheeler/documents/Electrodynamics/Miscellaneous\%20Essays/E\&M\%20in\%202\%20Dimensions.pdf}.

\bibitem{mcdonald}
K.T.~McDonald,
\emph{Electrodynamics in 1 and 2 spatial dimensions},
Joseph Henry Laboratories, Princeton University (2019),
\url{http://kirkmcd.princeton.edu/examples/2dem.pdf}.

\bibitem{boito}
D.~Boito, L.N.S.~de Andrade, G.~de Sousa, R.~Gama and C.Y.M.~London,
\emph{On Maxwell's electrodynamics in two spatial dimensions},
\textit{Rev. Bras. Ensino F\'{i}s.} \textbf{42} (2020) e20190323.

\bibitem{GBB}
\'{E}.~Goulart, E.~Bittencourt and E.O.S.~Brand\~ao,
\emph{Light propagation in $(2+1)$-dimensional electrodynamics: the case of linear constitutive laws},
\textit{Phys. Rev. A} \textbf{106} (2022) 063521
[\href{https://arxiv.org/abs/2209.00770}{arXiv:2209.00770}].

\bibitem{BBG}
E.~Bittencourt, E.O.S.~Brand\~ao and \'{E}.~Goulart,
\emph{Light propagation in $(2+1)$-dimensional electrodynamics: the case of nonlinear constitutive laws},
\href{https://arxiv.org/abs/2211.00477}{arXiv:2211.00477 [physics.optics]}.

\bibitem{Huygens}
S.~Pascazio, F.V.~Pepe and J.M.~P\'{e}rez-Pardo,
\emph{Huygens' principle and Dirac--Weyl equation},
\textit{Eur. Phys. J. Plus} \textbf{132} (2017) 287.

\bibitem{ref-JMP}
R.~Maggi, E.~Ercolessi, P.~Facchi, G.~Marmo, S.~Pascazio and F.V.~Pepe,
\emph{Dimensional reduction of electromagnetism},
\textit{J. Math. Phys.} \textbf{63} (2022) 022902
[\href{https://arxiv.org/abs/2111.10728}{arXiv:2111.10728}].

\bibitem{angelone23}
G.~Angelone, E.~Ercolessi, P.~Facchi, D.~Lonigro, R.~Maggi, G.~Marmo, S.~Pascazio and F.V.~Pepe,
\emph{Dimensional reduction of the Dirac theory},
\textit{J. Phys. A} \textbf{56} (2023) 065201
[\href{https://arxiv.org/abs/2211.08581}{arXiv:2211.08581}].

\bibitem{ForgacsManton} P. Forgács and N. S. Manton, \emph{Space-time symmetries in gauge theories,} Commun. Math. Phys. \textbf{72} (1980), 15

\bibitem{Manton} N. S. Manton, \emph{A new six-dimensional approach to the Weinberg–Salam model,} Nucl. Phys. B \textbf{158} (1979), 141
\bibitem{Hosotani1}
Y. Hosotani, \emph{Dynamical mass generation by compact extra dimensions,} Phys. Lett. B \textbf{126} (1983), 309
\bibitem{Hosotani2}
Y. Hosotani, \emph{Dynamics of non-integrable phases and gauge symmetry breaking,} Annals Phys. \textbf{190} (1989), 233

\bibitem{KapetanakisZoupanos}
D.~Kapetanakis and G.~Zoupanos,
\emph{Coset space dimensional reduction of gauge theories,}
Phys.\ Rept.\ \textbf{219} (1992) 1--76,
\href{https://doi.org/10.1016/0370-1573(92)90101-5}
{doi:10.1016/0370-1573(92)90101-5}.

\bibitem{ScherkSchwarz1979}
J.~Scherk and J.H.~Schwarz,
``How to Get Masses from Extra Dimensions,''
Nucl.\ Phys.\ B \textbf{153} (1979) 61--88,
doi:10.1016/0550-3213(79)90592-3.

\bibitem{CsakiGrojeanMurayama}
C.~Csáki, C.~Grojean and H.~Murayama,
\emph{Standard Model Higgs from higher dimensional gauge fields,}
Phys.\ Rev.\ D \textbf{67} (2003) 085012,
\href{https://doi.org/10.1103/PhysRevD.67.085012}
{doi:10.1103/PhysRevD.67.085012}
[\href{https://arxiv.org/abs/hep-ph/0210133}
{\texttt{hep-ph/0210133}}].

\bibitem{Serone}
M.~Serone,
\emph{Holographic methods and gauge-Higgs unification in flat extra
dimensions,}
New J.\ Phys.\ \textbf{12} (2010) 075013,
\href{https://doi.org/10.1088/1367-2630/12/7/075013}
{doi:10.1088/1367-2630/12/7/075013}
[\href{https://arxiv.org/abs/0909.5619}
{\texttt{arXiv:0909.5619 [hep-ph]}}].

\bibitem{HosotaniYamatsu}
Y.~Hosotani and N.~Yamatsu,
{Gauge-Higgs grand unification,}
Prog.\ Theor.\ Exp.\ Phys.\ \textbf{2015} (2015) 111B01,
\href{https://doi.org/10.1093/ptep/ptv153}
{doi:10.1093/ptep/ptv153}
[\href{https://arxiv.org/abs/1504.03817}
{\texttt{arXiv:1504.03817 [hep-ph]}}].

\bibitem{Ebert:2022ehb}
S.~Ebert, C.~Ferko, H.~Y.~Sun and Z.~Sun,
``$T\bar{T}$ in JT Gravity and BF Gauge Theory,''
SciPost Phys. \textbf{13} (2022) no.4, 096
doi:10.21468/SciPostPhys.13.4.096
[\href{https://arXiv:2205.07817}
{\texttt{arXiv:2205.07817 [hep-th]}}].

\bibitem{ChircoVacchianoVitale2026}
G.~Chirco, L.~Vacchiano and P.~Vitale,
``Dimensional reduction of $AdS_3$ Chern--Simons gravity:
Schwarzian and affine boundary theories,''
[\href{https://arXiv:2605.15293}
{\texttt{arXiv:2605.15293 [hep-th]}}].

\bibitem{Mertens:2018fds}
T.~G.~Mertens,
``The Schwarzian theory {\textemdash} origins,''
JHEP \textbf{05} (2018), 036
\href{https://doi:10.1007/JHEP05(2018)036}
[arXiv:1801.09605 [hep-th]].

\href{https://doi.org/10.1016/0370-1573(92)90101-5}
{doi:10.1016/0370-1573(92)90101-5}.




\bibitem{ScruccaSeroneSilvestrini}
C.~A.~Scrucca, M.~Serone and L.~Silvestrini,
\emph{Electroweak symmetry breaking and fermion masses from extra
dimensions,}
Nucl.\ Phys.\ B \textbf{669} (2003) 128--158,
\href{https://doi.org/10.1016/j.nuclphysb.2003.07.013}
{doi:10.1016/j.nuclphysb.2003.07.013}
[\href{https://arxiv.org/abs/hep-ph/0304220}
{\texttt{hep-ph/0304220}}].

\bibitem{Fujikawa}
K.~Fujikawa,
``Path-integral measure for gauge-invariant fermion theories,''
Phys.\ Rev.\ Lett.\ \textbf{42} (1979) 1195--1198,
\href{https://doi.org/10.1103/PhysRevLett.42.1195}
{doi:10.1103/PhysRevLett.42.1195}.

\bibitem{FujikawaSuzuki}
K.~Fujikawa and H.~Suzuki,
\textit{Path Integrals and Quantum Anomalies},
Oxford University Press, Oxford (2004).
\end{thebibliography}
\end{document}